\pdfoutput=1

\documentclass[11pt]{article}

\usepackage[]{ACL2023}

\usepackage{times}
\usepackage{latexsym}
\usepackage{listings}
\usepackage{booktabs}       
\usepackage{amsfonts}       
\usepackage{nicefrac}       
\usepackage{microtype}      
\usepackage{multirow}

\usepackage{float}
\usepackage{tabulary}
\usepackage{tabularx}
\usepackage{amsmath}
\usepackage{listings}
\usepackage{algorithm}
\usepackage[normalem]{ulem}
\usepackage{algorithmic}
\usepackage{graphbox}
\usepackage{subcaption}
\usepackage{graphicx}

\usepackage[T1]{fontenc}

\usepackage[utf8]{inputenc}

\usepackage{microtype}

\usepackage{inconsolata}

\title{CiteFix: Enhancing RAG Accuracy Through Post-Processing Citation Correction}

\author{Harsh Maheshwari \\
  \texttt{mahhars@amazon.com} \\\And
  Srikanth Tenneti \\
  \texttt{stenneti@amazon.com} \\ \And 
  Alwarappan Nakkiran \\
  \texttt{nakkiran@amazon.com}}

\begin{document}
\maketitle
\begin{abstract}
Retrieval Augmented Generation (RAG) has emerged as a powerful application of Large Language Models (LLMs), revolutionizing information search and consumption. RAG systems combine traditional search capabilities with LLMs to generate comprehensive answers to user queries, ideally with accurate citations. However, in our experience of developing a RAG product, LLMs often struggle with source attribution, aligning with other industry studies reporting citation accuracy rates of only about 74\% for popular generative search engines. To address this, we present efficient post-processing algorithms to improve citation accuracy in LLM-generated responses, with minimal impact on latency and cost. Our approaches cross-check generated citations against retrieved articles using methods including keyword + semantic matching, fine tuned model with BERTScore, and a lightweight LLM-based technique. Our experimental results demonstrate a relative improvement of 15.46\% in the overall accuracy metrics of our RAG system. This significant enhancement potentially enables a shift from our current larger language model to a relatively smaller model that is approximately 12x more cost-effective and 3x faster in inference time, while maintaining comparable performance. This research contributes to enhancing the reliability and trustworthiness of AI-generated content in information retrieval and summarization tasks which is critical to gain customer trust especially in commercial products.
\end{abstract}

\section{Introduction} \label{sec_intro}

Recent advancements in AI infrastructure and methodologies have enabled training Large Language Models (LLMs) over internet-scale data. These models demonstrate impressive competence in answering a wide range of general queries. However, when applied to specialized domains such as addressing questions based on internal company documents, off-the-shelf LLMs exhibit significant limitations. They often lack access to latest information, have difficulty interpreting domain specific language, struggle with source attribution, are prone to hallucinations \cite{ji2023survey}, and are prone to overly broad responses.

To overcome these challenges, two broad strategies have emerged. The first involves fine-tuning LLMs on domain-specific data. However, this approach is not only resource-intensive and requires frequent updates, but also risks unintended consequences such as catastrophic forgetting, where the model loses previously acquired general knowledge, thereby increasing the overall system complexity. The second, often more practical method is Retrieval-Augmented Generation (RAG). RAG is a process that combines information retrieval with text generation. It typically involves the following steps: (1) indexing a knowledge base of relevant information, (2) using a retrieval system to find content specifically relevant to a given user query, (3) providing the user query and the retrieved content to an LLM, instructing it to generate a response based on the retrieved content. RAG offers numerous benefits, including real-time access to up-to-date information, improved token generation \cite{khandelwal2019generalization}, reduced hallucinations, better source attribution \cite{gao2023enabling, hsu2024calm} and overall superior response generation  \cite{shuster2021retrieval, bechard2024reducing}. Additionally, RAG tends to be more cost-effective and transparent than full model fine-tuning. Examples of RAG-based products include Perplexity.ai \cite{PerplexityAI}, bing search, GPT Search etc.

Despite enabling a novel information retrieval experience for users, RAG systems today face key limitations. Table ~\ref{tab:basellm} illustrates results of a Subject Matter Expert based auditing of a RAG based system. Shown is a metric "Relative Mean Question Level Accuracy", which captures relevancy of cited chunks, correctness and completeness of the answer(Sec.~\ref{sec:Metric}), relative to Model C\footnote{Model names are anonymized following standard practice for proprietary/pre-release models. Publicly available models retain their original names. Model A, Model B and Model C are sufficiently large and powerful language model. With number of parameters in decreasing order for A, B and C. Model B however is the model trained on latest data with better methodologies} accuracy. A prevalent form of error that contributes to lower performance is that of unverifiable facts in LLMs' responses. Unverifiable facts are the facts in LLM response which cannot be validated by cited reference. In our analysis for Model C, notably around 80\% of these unverifiable facts were not pure hallucinations, but rather errors in the model's ability to cite the correct reference from which it generated the given factual point. These observations align with industry studies \cite{liu2023evaluatingverifiabilitygenerativesearch} reporting citation accuracy rates of only about 74\% for popular generative search engines. Incorrect citations not only reduce the actionability of the responses, but also dent customer trust, especially for commercial products. This paper focuses on this issue and proposes methods to address it.

\begin{figure}
    \centering
    \includegraphics[width=0.75\linewidth]{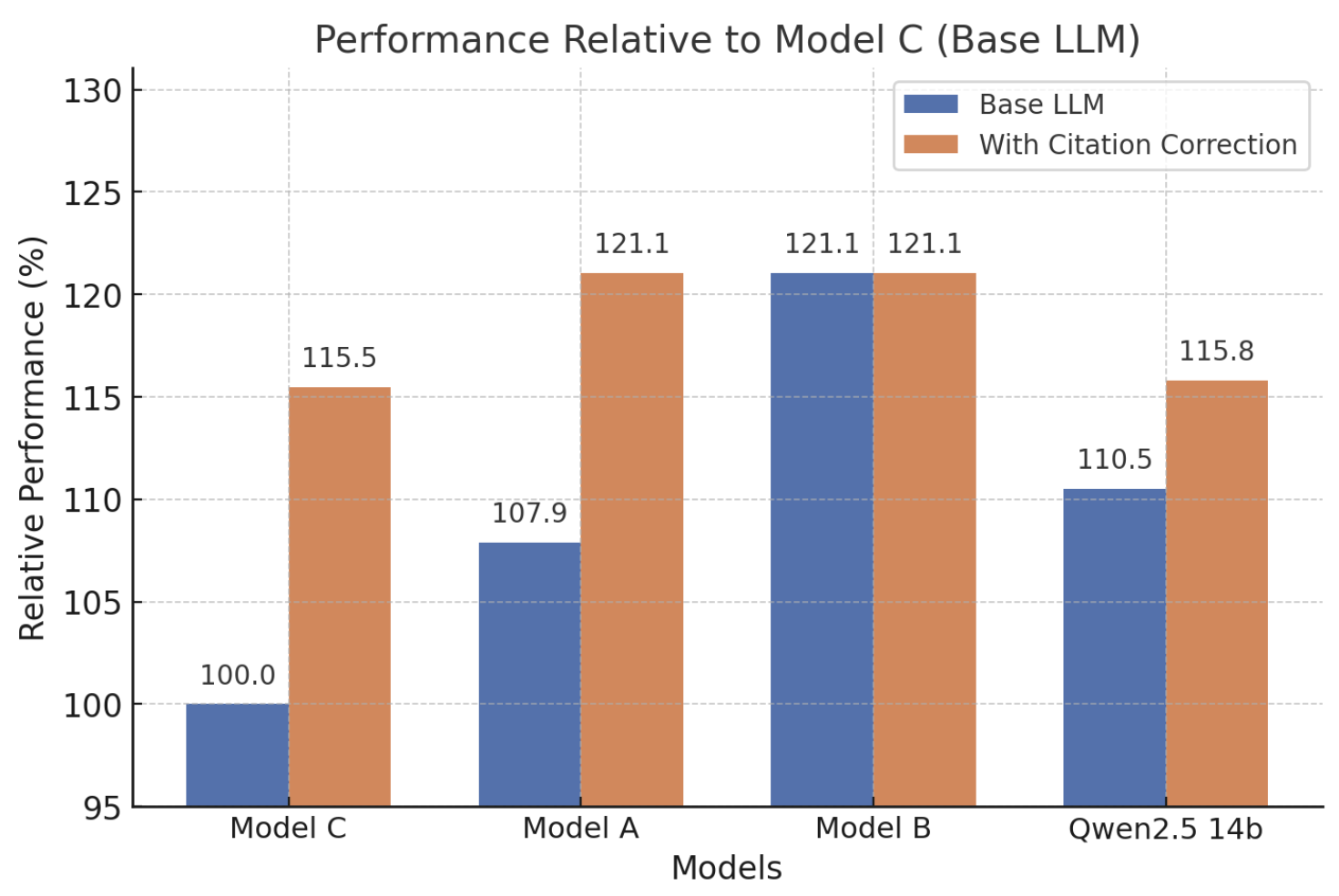}
    \caption{Improvements in RAG accuracy for various LLMs after employing our proposed citation correction methods. Results are shown as percentage improvements in Mean Question Level Accuracy(MQLA) over Model C baseline performance without citation correction. MQLA is a metric designed to capture relevancy, correctness and completeness (see Sec.~\ref{sec:Metric}).}
    \label{fig:performance_image}
\end{figure}

\begin{table}[t]
\centering
\resizebox{1.1\columnwidth}{!}{
\begin{tabular}{llc|ccc}
\hline


\textbf{Model} & \textbf{Cents per 1K}     & \textbf{Relative Mean Question}  & \textbf{\% of factual}       & \textbf{\% of factual points} & \textbf{\% of factual points} \\
         & \textbf{O/P tokens} & \textbf{Level Accuracy} & \textbf{points unverifiable} & \textbf{incorrectly cited}
& \textbf{purely hallucinated} \\
\hline
Model A  & +1100\%         & +7.9\% (+12\%) & Base (Base) & 90.8\% (65\%)  & 9.1\% (35\%)\\
Model B & +220\%         & +21.1\% (+21.1\%) & Base (Base)  & 66.6\% (66.6\%)  & 34.4\% (-33.4\%)\\
Model C & Base       & Base (+15.4\%)  & Base (Base) & 80.6\% (33.3\%)  &  19.4\%(-66.6\%)\\
Qwen 14-B & open source & +10.5\% (+15.8\%)  & Base (Base)  & 76.2\% (70.8\%)  & -13.8\% (29.2\%)\\
Qwen 2-B  & open source & -39\% (NA)        & NA & NA      & NA   \\
\hline \\
\end{tabular}
}

\caption{\textit{Motivating the need for CiteFix:} This table shows the prevalence of incorrect citations across LLMs and our method's impact. Model C is the baseline for cost and accuracy columns. For the last three columns, the baseline is each model's total percentage of unverifiable factual points. Numbers outside (inside) parentheses show performance before (after) CiteFix. Initially, incorrect citations significantly outnumber hallucinations. CiteFix balances this ratio and in absolute terms it drastically reduces incorrect citations. Qwen 2-B was excluded from detailed audit due to inconsistent citation generation.}

\label{tab:basellm}
\end{table}

While previous studies have explored attributable text generation (\cite{nakano2022webgptbrowserassistedquestionansweringhuman}; \cite{gao-etal-2023-enabling}) and simple prompting techniques for citation incorporation (\cite{malaviya2024expertqaexpertcuratedquestionsattributed, sun2024verifiabletextgenerationevolving, li2024verifiablegenerationbenchmarkknowledgeaware}), systematic evaluations reveal significant performance gaps \cite{gao-etal-2023-enabling}. Recent work \cite{huang2024traininglanguagemodelsgenerate} has only scratched the surface by demonstrating attribution quality degradation from ChatGPT to Llama 2 7B, leaving a critical need for deeper analysis and practical solutions.

This paper offers two contributions:

\begin{enumerate}
    \item Demonstrating the existence and extent of the incorrect citations issue across multiple LLMs, and highlighting the need to address the same.     
    \item Proposing six computationally light weight methods to address this issue, ranging from simple heuristic methods to more sophisticated learning-based solutions. Through extensive experimentation, we show that different citation correction approaches may be optimal for different LLMs - for instance, hybrid (lexical + semantic) matching works best with Model A, while fine-tuned BERTScore performs better with Model B. We provide detailed comparisons of their effectiveness and practical applicability. As seen in Fig ~\ref{fig:performance_image} and Table~\ref{tab:basellm}, our method resulted in an improvement of upto 15.46\% relative improvement in accuracy when tested across four different LLMs.
\end{enumerate}

Through this work, we aim to not only advance the understanding of citation accuracy challenges in LLMs, but also provide practical low cost solutions for improving attribution in real-world applications. Sec.~\ref{sec_rel_work} presents an overview of related work. Sec.~\ref{sec_method} details our proposed algorithms. Sec.~\ref{sec_results} presents evaluation results. Sec.~\ref{sec_conclude} concludes, along with a discussion of the limitations of our work and plans for addressing them going forward.

\section{Related Work}
\label{sec_rel_work}

Accurate attribution of information to sources remains a critical challenge in building trustworthy AI systems, particularly for Large Language Models (LLMs) and Retrieval-Augmented Generation (RAG) systems. The challenge of accurate attribution in AI-generated content has been approached from multiple angles in the literature. Some researchers have focused on developing models specifically designed for attributable text generation \cite{nakano2022webgptbrowserassistedquestionansweringhuman}, while others have explored the effectiveness of prompt engineering techniques for citation accuracy \cite{malaviya2024expertqaexpertcuratedquestionsattributed, li2024verifiablegenerationbenchmarkknowledgeaware}. However, a comprehensive study \cite{gao-etal-2023-enabling} has highlighted that significant challenges remain, particularly in maintaining consistent attribution accuracy across different types of queries and document structures. These findings underscore the need for more robust and versatile approaches to citation/attribution in AI systems.

Recent work has focused on the automatic evaluation of attribution by LLMs \cite{yue2023automaticevaluationattributionlarge} and factual entailment for hallucination detection \cite{rawte2024factoidfactualentailmenthallucination}, primarily assessing whether generated content is present in cited references. However, there is a notable gap in research specifically addressing citation correction.

Many existing methods, including those fine-tuning T5 models \cite{gao2023enablinglargelanguagemodels, song2024measuringenhancingtrustworthinessllms, honovich2022truereevaluatingfactualconsistency}, are limited by context lengths of around 512 tokens. This constraint poses significant challenges when dealing with longer documents or multiple sources, which is often the case in practical RAG systems. Our proposed solution for citation correction is designed to handle larger context lengths, addressing a critical limitation in current approaches.

Furthermore, our research distinguishes itself by focusing on not just detecting citation errors but actively working towards correcting them. This shift from identification to correction represents a significant step forward in improving the usefulness of AI-generated content in RAG systems. We introduce a range of citation correction methods, including lexical matching, hybrid (lexical + semantic) approaches, and lightweight LLM-based attribution. One method builds on BERT Score \cite{zhang2020bertscoreevaluatingtextgeneration}, leveraging pre-trained contextual embeddings from BERT \cite{devlin2019bertpretrainingdeepbidirectional}. Initial experiments with an off-the-shelf model \cite{beltagy2020longformerlongdocumenttransformer} showed improvements, but fine-tuning on in-domain data yielded better results. This led us to explore ColBERT \cite{khattab2020colbertefficienteffectivepassage}, a neural retrieval model designed for fine-grained contextual late interaction. By combining BERT Score’s semantic similarity assessment with ColBERT’s fine-tuning capabilities, we developed a more robust and accurate citation correction method, which we detail in the next section. We detail these methods in the next section.

\section{Proposed Methodology} 
\label{sec_method}
Our goal is to improve the overall citation accuracy, while having minimal impact on latency and costs. Towards this, we propose a suite of algorithms that leverage various techniques, ranging from simple heuristics to sophisticated machine learning models. Our algorithms are streaming-compatible post-processing techniques, meaning that they operate on an LLM's response as it is being generated.

\begin{figure}
    \centering
    \includegraphics[width=0.90\linewidth]{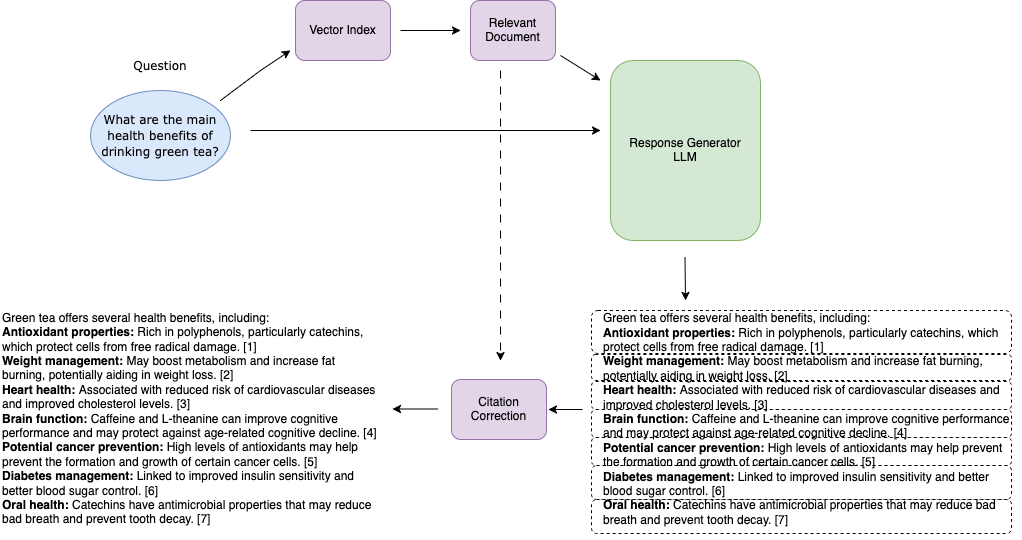}
    \caption{Overview of the workflow of the proposed methods using a sample question. Once the RAG system's response generating LLM generates an answer, we split the answer into distinct factual points (shown in dotted lines above). For each factual point, we use its similarity scores with the retrieved documents to detect citation errors and correct them. See Section ~\ref{sec_method} for details. Question used is for illustration purpose only}
    \label{fig:citation_workflow}
\end{figure}

The general framework of our proposed methods is depicted in Figure ~\ref{fig:citation_workflow}. We will now go into its details. Let us denote the query that the user asks the RAG system as $q$. Let the set of documents retrieved by the Retriever module in RAG be $\{\hat{x}_{i}\}_{i=0}^{R-1}$. Let $A$ denote the answer generated by the LLM. Our algorithms involve the following steps:
\begin{enumerate}
    \item We first split the LLM's response $A$ into distinct "factual points" $\{x_{i}\}_{i=0}^{L-1}$. A factual point is defined as a section within $A$ that the LLM attributes to a particular set of retrieved documents via citations. In our use case, the LLMs were instructed to include citations at the end of each factual statement in their response. We use simple regular expressions to segment the LLM's response into "factual points", delimited by citations. See Fig.~\ref{fig:citation_workflow} for example.

    \item Let $C_i$ be the number of citations in the LLM's generated response $A$ for the factual point $x_i$. Our algorithms will estimate the "corrected citations" to be the top $C_i$ retrieved documents among $\{\hat{x}_{i}\}_{i=0}^{R-1}$ that maximize the following similarity metric with the factual point $x_i$:
    \begin{equation}
    \label{eq:sim}
        s_{ij} = f(x_i, \hat{x}_{j})
    \end{equation}
\end{enumerate}

In the next sections, we will discuss various choices for the function $f$ in Eq.~\ref{eq:sim}. We will use the following notation: Let us denote each factual point $x_i$ as list of its individual tokens $t_{ij}$. Namely, $x_i = [t_{i0}, t_{i1}, \ldots , t_{ik}]$. Let us also denote each retrieved document $\hat{x}_{i}$ as a list of its tokens $\hat{x}_{i} = [\hat{t}_{i0}, \hat{t}_{i1}, \ldots ,\hat{t}_{il}]$.

\subsection{Keyword based matching}

We define $f$ in Eq.~\ref{eq:sim} as the size of the intersection between the tokens in $x_i$ and $\hat{x}_{j}$. We also tried a term-frequency (TF) by inverse-document-frequency type of scoring, such as done in traditional document ranking \cite{rousseau2013composition, trotman2014improvements}, but it did not yield good results. We noticed regular IDF being particularly noisy with domain specific keywords such as "yield" which have different meaning in agriculture and financial context or "drill" which have different meaning in mining and military context etc.

\subsection{Keyword + Semantic Context based matching}

In this approach, we combine the above keyword match score with a mild contribution from the semantic similarity between the user query $q$ and the retrieved document $\hat{x}_i$. The motivation is to mildly prefer retrievals that are more relevant to the user query:
\begin{equation}
f(x_i, \hat{x}_{j}) = \lambda . f_{keyword}(x_i, \hat{x}_{j}) + (1 - \lambda) . r(q, \hat{x}_j)
\end{equation}
Where $f_{keyword}(x_i, \hat{x}_{j})$ is the keyword based matching score and $r(q, \hat{x}_j)$ is the retrieval score for document $\hat{x}_j$ given query $q$. We empirically found $\lambda = 0.8$ to perform well in our experiments.

\subsection{BERT Score}

In the previously discussed approaches, contextual meaning of the words in $x_i$ and $\hat{x}_j$ was not fully utilized. They also do not differentiate between cases where word matches occur in close proximity within the reference versus where they are scattered across unrelated positions. Additionally, keyword-based methods struggle to handle scenarios where the language model or response generator paraphrases the words, as these methods rely on exact word matches.

BERT Score \cite{zhang2020bertscoreevaluatingtextgeneration} addresses these limitations by leveraging contextual embeddings to represent the tokens in the factual point $x_i$ and the reference $\hat{x}_{j}$. These embeddings are generated using the LongFormer model \cite{beltagy2020longformerlongdocumenttransformer}, which incorporates bi-directional attention to capture not only the token but also its surrounding context.

Once the embeddings are computed, the similarity between the factual point and a retrieved document is calculated as follows: For each token in the factual point $x_i$, we compute its maximum similarity among all tokens in the retrieved document. The mean of these maximum similarity scores among all tokens in $x_i$ is used as  the final score in Eq~\ref{eq:sim}:

\begin{equation}
\label{bert_score}
f(x_i, \hat{x}_{j}) = \frac{1}{|x_i|} \sum_{t_{il} \in x_i} \max_{\hat{t}_{jk} \in \hat{x}_j} e(t_{il})^\top e(\hat{t}_{jk})
\end{equation}
where $e(t)$ denotes the embedding of a token $t$. 

\subsection{Fine-tuned Model with Bert Score}
While off-the-shelf  BERTScore models provide a good starting point for incorporating contextual similarity into the citation correction process, we hypothesize that fine-tuning these models specifically for this task on an in-domain dataset can further improve their performance. The key limitation of the off-the-shelf models is that they are not explicitly trained to capture the nuances of citation attribution \& factual entailment. Our methodology is motivated by ColBERT \cite{khattab2020colbertefficienteffectivepassage}.

During training, the input to the model is a factual point ($x$), a positive reference ($\hat{x}{+}$) that validates the point, and a negative reference ($\hat{x}{-}$) that does not validate the factual point. BERTScore for the factual point, calculated using Eq.~\ref{bert_score}, is maximized for the positive reference compared to the negative reference. We used cross-entropy loss to increase the score with the positive reference compared to the negative reference.


\textbf{Dataset Preparation}: To train the model, we need factual points, and corresponding positive and negative references. We employed an LLM for this, using two strategies: First, for each document in the corpus, we determine the $n^{th}$ most similar document using \cite{titanv2}. We then prompt LLM to provide a factual point present in the former document, but not in the latter. By varying $n \in \{14, 11, 8, 5, 4, 3\}$, we get progressively hard positive and negative pairs for training. Secondly, for a list of questions, we generate answers from our RAG-based system. For each factual point present in the answer and for each retrieved document, we employ an LLM to check for if the former is grounded in the latter. We then use this information to create multiple pairs of positive-negative for a given factual point. This allows us to tune the model specifically for the citations issue for the specific LLM used within the RAG system.

\subsection{LLM Based Matching}
An alternative approach for citation correction is to employ an LLM directly. Table~\ref{tab:basellm} presents results using our best-performing prompt instructions for citation-aware response generation. Here, we explore a secondary LLM that identifies the most relevant reference for each factual point.

To balance accuracy with efficiency, we use a simple prompt that requests only the reference number, avoiding complex techniques like Chain of Thought (CoT) \cite{wei2023chainofthoughtpromptingelicitsreasoning}, which would increase token usage, latency, and cost. This approach leverages the LLM’s ability to capture contextual and semantic nuances beyond keyword-based or rule-based methods, enabling adaptability across domains without explicit rule-crafting or fine-tuning.

However, the effectiveness of this method depends on the LLM’s quality, training data, and prompt design. Additionally, processing each factual point individually introduces computational overhead, requiring a careful trade-off between cost, latency, and accuracy.

\subsection{Reusing Attention Maps of the Base LLM} \label{sec:attn}
The main idea here is, can we look at the attention maps of the response generating LLM itself to check which retrieved documents were used in generating each factual point in the response. We did not have enough time to fully experiment with this idea, but in  Appendix ~\ref{app:attention_map}, we show a simple proof of concept that demonstrates this idea. We will explore this further in our future work.

\section{Results} \label{sec_results}
In this section, we will present evaluation results of all the proposed methods on top of RAG based system. The evaluations were done by human auditors, who have prior knowledge on the topic for which RAG is used.

\subsection{Metrics}
\label{sec:Metric}
We developed the following metrics to evaluate RAG system performance. The uber level metric we track is called "Mean Question-Level Accuracy" (MQLA). It combines the following:
\begin{itemize}
    \item \textbf{Relevancy URL}: Checks if the set of citations referenced to by the LLM are relevant to the question. Calculated as the fraction of cited URLs that are relevant.
    \item \textbf{Relevancy Keywords}: Checks if keywords in the LLM's response are relevant to the question. Calculated as the ratio of keywords which are relevant by the total number of keywords present in the query. The keywords in the response are identified by humans.
    \item \textbf{Relevancy Facts}: Checks if facts present in the LLM's response are relevant to the question. Calculated as the ratio of facts which are relevant to query by the total number of facts present in the response. The facts in the response are identified by humans.
    \item \textbf{Correctness}: Checks if the facts present in the LLM's response can be verified in the citations provided. Calculated as the ratio of number of facts supported by cited references and the total number of facts. \textbf{Note}: The facts not supported by cited referenced can be divided into two categories 1) Hallucinated facts and 2) Incorrectly cited facts, based on whether the fact was present in any of the retrieved documents or not.
    \item \textbf{Completeness}: Checks if all aspects (possible sub-questions) of the original questions are addressed in the response. The possible sub-questions are identified by the humans.
\end{itemize}
We calculate MQLA as described in Algorithm ~\ref{algo:question_accuracy}.
 
\begin{algorithm}
\caption{Mean Question Level Accuracy}
\label{algo:question_accuracy}
\begin{algorithmic}[1]
\STATE Initialize totalAccuracy$ = 0$, n $=$ number of questions 
\FOR{q in questions}
\STATE Initialize accuracy$ = 0$
\IF{all(relevancyUrl, relevancyKeyword, relevancyFacts, correctness, completeness $\geq 0.8$) and hallucinatedFacts $\leq 1$} 
    \STATE accuracy$ = 1$
    \ENDIF 
    \STATE totalAccuracy $+=$ accuracy \ENDFOR 
\STATE meanAccuracy $=$ totalAccuracy / n 
\RETURN meanAccuracy
\end{algorithmic}
\end{algorithm}

\subsection{Comparing Different Citation Correction Methods} \label{sec:haiku_comp}

\begin{table*}[t]
\caption{Comparing Citation Correction Methods. All columns except p90 latency show relative performance}
\label{tab:compare_haiku3}
\centering
\resizebox{0.82\linewidth}{!}{%
\begin{tabular}{lccccc}
\hline
\multicolumn{1}{c}{\textbf{\begin{tabular}[c]{@{}c@{}}Citation Correction \\ Method\end{tabular}}} & \textbf{\begin{tabular}[c]{@{}c@{}}Response Generating\\  LLM\end{tabular}} & \textbf{\begin{tabular}[c]{@{}c@{}} Mean Question \\ Level Accuracy \end{tabular}} & \textbf{Relevancy URL} & \textbf{\begin{tabular}[c]{@{}c@{}}\% of Facts \\ Correctly Cited\end{tabular}} & \textbf{\begin{tabular}[c]{@{}c@{}} p90 latency per \\ factual point (in sec) \end{tabular}} \\ \hline
None & Model C & Base & Base & Base & - \\  \hline
Keyword & Model C & +12.7\% & -0.9\% & +12\% & 0.014 \\ \hline
Keyword + Semantic Context & Model C & +15.5\% & -0.9\% & +13.6\% & 0.015\\ \hline

BERT Score & Model C & +2.6\% & -1\% & +3.2\% & 0.389\\ \hline
Finetuned BERT Score & Model C & \bf{+15.8\%} & +1.5\% & \bf{+13.7\%} & 0.389 \\ \hline
LLM Based Matching (Model C) & Model C & +1.9\% & +0.9\% & +7\% & 1.586 \\ \hline
None (Baseline) & Model A & +7.8\% & \bf{+2\%} & +5.4\% & - \\ \hline
\end{tabular}%
}
\end{table*}

In Table~\ref{tab:compare_haiku3}, we compare different citation correction algorithms proposed in this paper on Model C's responses. We used a set of 50 representative questions for evaluation, incurring an audit time of 2.5 days by 2 humans per row of Table~\ref{tab:compare_haiku3}. The table includes p90 latency per factual point for each citation correction method, which adds negligible overhead (except LLM method) to our system's time to first token p90 latency. The latency is computed on g5.4xlarge instance. Results for Model A, a model that is 12x more expensive and about 3x slower are also shown for reference. The impact of our techniques Keyword + Semantic Context based and Fine-tuned BERT Score is evident, taking Model C's MQLA higher than Model A. 

\subsection{Evaluating Impact Across Different LLMs}

In Table~\ref{tab:LLMCompare}, we evaluated the two best performing citation correction methods from Table~\ref{tab:compare_haiku3} for four different LLMs (using the same dataset as in Sec.~\ref{sec:haiku_comp}). Interestingly, different LLMs may pair optimally with different citation correction strategies. The impact of our methods is strongly evident for Model C, Model A and Qwen 2.5 14-B. Model B seems to be inherently much better at citations, but we see some mild improvements in the relevancy of cited URLs when paired with our fine-tuned BERT Score method. These results demonstrate potentially wide applicability of our proposed methods.

\begin{table}[]
\caption{This table shows the effectiveness of our two best citation correction approaches with various LLMs. KSC represents Keyword+Semantic context and FBS represents Finetuned BERT Score}
\label{tab:LLMCompare}
\centering
\resizebox{0.85\linewidth}{!}{%
\begin{tabular}{lcccc}
\hline
\multicolumn{1}{c}{\textbf{\begin{tabular}[c]{@{}c@{}}Response \\ Generator \end{tabular}}} & \textbf{\begin{tabular}[c]{@{}c@{}}Citation \\ Corrector \end{tabular}} & \textbf{MQLA} & \textbf{\begin{tabular}[c]{@{}c@{}}Relevancy \\ URL \end{tabular}} & \textbf{\begin{tabular}[c]{@{}c@{}}\% of facts \\ Correctly Cited\end{tabular}} \\ \hline
Model C & None & base & base & base \\
Model C & KSC & +15.5\% & -0.9\% & +13.6\% \\
Model C & \textbf{FBS}  & \textbf{+15.8\%}  & \textbf{+1.5\%} & \textbf{+13.7\%} \\ \hline
Model B & \textbf{None} & \textbf{+21\%} & +1.5\% & +14.9\% \\
Model B & KSC & +10.5\% & +1.5\% & +10.7\% \\
Model B & \textbf{FBS} & \textbf{+21\%} & \textbf{+2\%} & \textbf{+15\%} \\ \hline
Model A & None & +7.9\% & \textbf{+2\%} & +5.4\% \\
Model A & \textbf{KSC} & \textbf{+21\%} & -1.3\% & \textbf{+16\%} \\
Model A & FBS & +10.5\% & \textbf{+2\%} & +9.8\% \\ \hline
Qwen 2.5 14b & None & +10.5\% & \textbf{+2\%} & +8.4\% \\
Qwen 2.5 14b & \textbf{KSC} & \textbf{15.8\%} & \textbf{+2\%} & \textbf{+9.7\%} \\
Qwen 2.5 14b & FBS & 13.1\% & +1.3\% & +8.7\% \\ \hline
\end{tabular}%
}
\end{table}

\section{Conclusion} \label{sec_conclude}

This paper addresses the critical challenge of citation accuracy in RAG systems, demonstrating its impact across multiple LLMs and its effect on AI-generated content trustworthiness. Our key contribution is the development of efficient post-processing algorithms for citation correction, improving relative accuracy by up to 15.46\% while maintaining minimal computational overhead. Notably, we found that optimal citation correction methods vary across LLMs, emphasizing the importance of model-specific approach selection.

Our findings, while promising, represent early steps in addressing this challenge. Future research areas include exploring attention-map-based methods for more precise attributions and developing sophisticated dataset preparation techniques. While newer LLMs (Model B) have improved citation accuracy, attribution issues persist to a lesser extent, suggesting the need for more sophisticated correction algorithms. Additionally, our framework's ability to establish relationships between factual points and source documents opens up interesting applications, such as determining appropriate contexts for content insertion (e.g., advertisement placement) based on document similarity and factual relevance.

\bibliography{anthology,custom}
\bibliographystyle{acl_natbib}

\newpage
\section{Appendix}

\subsection{Using Attention map for attribution}
\label{app:attention_map}
In a RAG based system, the response generating LLM is given a set of relevant document in response to a user query. It then understands information from these different documents to answer the question at hand. Here, we want to explore if can we leverage attention scores within the LLM to understand which document in the prompt it is focusing on while generating a particular fact in its response. We did a small \textit{toy experiment} with Qwen 2.5B - 2B to test the same. We use the below prompt:

\begin{figure}
    \centering
    \includegraphics[width=0.8\linewidth]{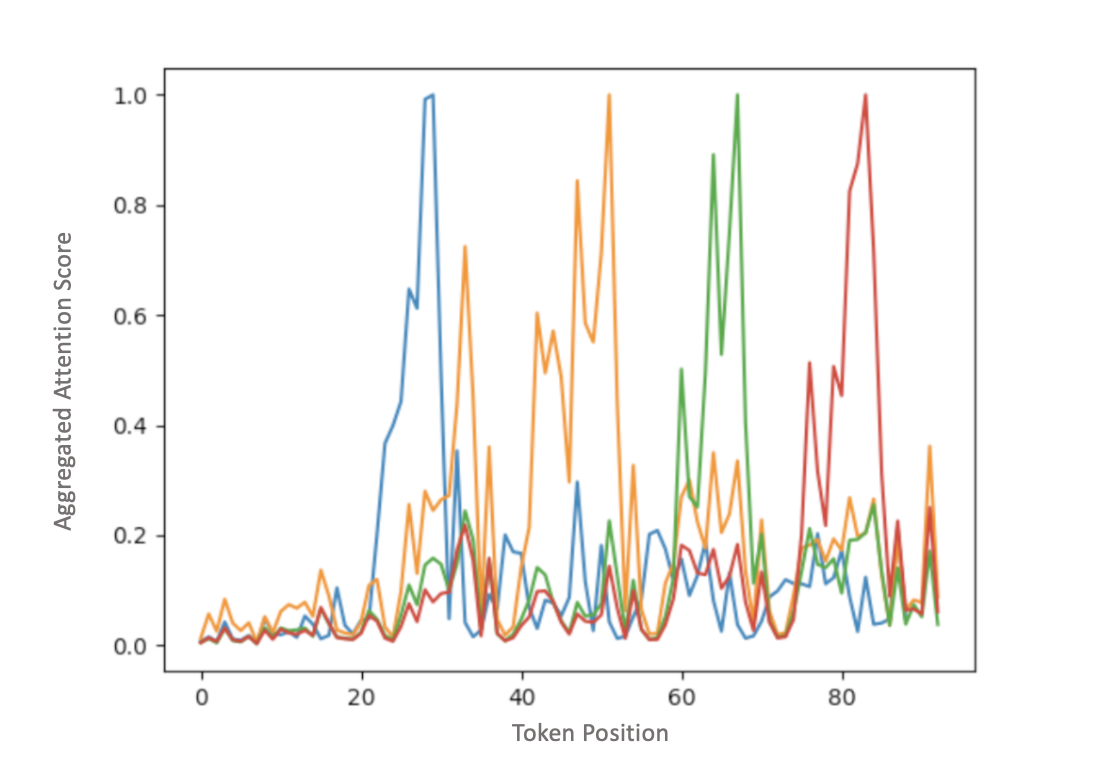}
    \caption{Visualisation of Attention Score. See Appendix~\ref{app:attention_map} for details.} 
    \label{fig:attention_score}
\end{figure}

\begin{lstlisting}
Hi, you are an assistant who has access to the following <documents> about cricket.
Please answer the <user query> at the end using only the information provided in
the <documents>. Do not output any information not contained in the <documents>.
Do not output any information that is not relevant to answering the <user query>.
If the <user query> cannot be answered with the given <documents>, please say so.

<documents>

<doc> Axx is a tall batsman. </doc>

<doc> Byy can bat with a broken bat as well. </doc>

<doc> Czz is a very funny umpire. </doc>

<doc> Dii is a fast bowler from Mumbai. </doc>

</documents>

<user query>
QUESTION
</user query>
\end{lstlisting}

We asked the following questions to the LLM: 
\begin{itemize}
    \item Name a batsman who is not particularly short
    \item Name a batsman who can bat with a damaged bat
    \item Name an umpire who makes people smile
    \item Who is a player from Mumbai?
\end{itemize}
and visualised the attention scores in ~\ref{fig:attention_score} (Blue, Orange, Green and Red lines for the above four questions respectively). The x-axis in the figure is the token position within the prompt. The y-axis is the sum of the attentions scores for all tokens in the output, across all layers of the LLM at that particular input token location. A higher value of this sum at a particular location of the input token indicates that that input token was taken into account by the LLM in generating the response.

You will see that for first question the peak of attention score is before the second question which is in line with where the necessary information is present in the prompt. Likewise, the peak of attention for second question is before the third one, and so on. This small proof of concept shows that we may be able to leverage the LLM's internal attention maps to correct citations.

\end{document}